\documentstyle[preprint,eqsecnum,aps]{revtex}
\tightenlines
\def\Cha{1}
\def\ScHe{2}
\def\TTRD{3}
\def\Dra{4}
\def\DKN{5}
\def\DaMu{6}
\def\Hen{7}
\def\LHR{8}
\def\BaTr{9}
\begin{document}
\draft
\title{Threshold amplitudes for transition to turbulence
in a pipe} 
\author{Lloyd N. Trefethen\cite{LNTbyline}}
\address{Oxford University Computing Laboratory, Wolfson Bldg., Parks Road, Oxford OX1 3QD, UK}
\author{S. J. Chapman\cite{SJCbyline}}
\address{OCIAM, Mathematical Institute, 24--29 St. Giles', Oxford OX1 3LB, UK}
\author{Dan S. Henningson\cite{DSHbyline}}
\address{Dept.\ of Mechanics, Royal Institute of Technology, S-10044 Stockholm, Sweden}
\author{\'Alvaro Meseguer\cite{AMbyline}}
\address{Oxford University Computing Laboratory, Wolfson Bldg., Parks Road, Oxford OX1 3QD, UK}
\author{Tom Mullin\cite{TMbyline}}
\address{Dept.\ of Physics and Astronomy, University of Manchester, Manchester M13 9PL, UK}
\author{F. T. M. Nieuwstadt\cite{FTMNbyline}}
\address{J. M. Burgers Centre, Delft U. of Technology, Rotterdamseweg 145, 2628 AL Delft, Netherlands}
\date{\today}
\maketitle
\begin{abstract}Although flow in a circular pipe is stable to
infinitesimal perturbations, it can be excited to turbulence
by finite perturbations whose minimal amplitude shrinks as
$R\to\infty$ ($R$ = Reynolds number).
Laboratory experiments have appeared to disagree with
one another and with theoretical predictions about
the dependence of this minimal amplitude on $R$, with published results
ranging approximately from $R^{-1/4}$ to $R^{-3/2}$.\ \ Here it is
shown that these discrepancies can be explained by the
use of different definitions of amplitude by different authors.
An attempt is made to convert the existing results to a uniform
definition of amplitude, the nondimensionalized $L^2$
definition common in the theoretical literature.
Although subtleties in the physics raise some questions,
agreement appears to be reached on a minimal amplitude
that scales as $R^{-3/2 \pm 0.3}$.
\end{abstract}

\narrowtext

\section{Introduction}

Laminar incompressible flow in an infinite circular pipe is mathematically stable,
but in practice, pipe flows invariably undergo transition to
turbulence if the Reynolds number $R$ is high.
It is generally accepted that an
explanation for this phenomenon is that although the laminar state
is stable to infinitesimal perturbations of the velocity field, 
certain small finite amplitude perturbations are enough
to excite transition for large $R$.\ \ A natural question
is, if $\epsilon = \epsilon(R)$ denotes the minimal 
amplitude of all perturbations that may excite transition, and if
$\epsilon$ scales with $R$ according to
\begin{equation}
\epsilon = O(R^{\kern 1pt\gamma})
\label{eq1}
\end{equation}
as $R\to \infty$, then what is the exponent $\gamma\,$?\ \ A value of $\gamma$
substantially below zero would correspond to a sensitivity of the laminar
flow that increases rapidly with $R$.

We six, coming from diverse backgrounds in applied mathematics, scientific computing,
and laboratory experimentation, have all been interested in (\ref{eq1}), but
the exponents $\gamma$ that our different lines of research have
suggested have varied by as much as a factor of six, from
$\approx \!-1/4$ to $-3/2$.  In
discussions at the ERCOFTAC Workshop on Subcritical Transition
in Delft in October, 1999, it
became clear that we have been using inconsistent definitions of the amplitude
of a velocity perturbation. 
Without a consistent definition, (\ref{eq1}) of course has little meaning.
The purpose of this note is to attempt to cast
our various results in terms of a single definition of amplitude.
The definition we shall use is essentially the one employed previously
by Chapman\cite{Cha}, Schmid and Henningson\cite{ScHe}, Trefethen,
et al.\cite{TTRD}, and
others, and we shall call it the {\em $L^2$ amplitude}.\ \ We do
not argue that this definition is more or less appropriate physically than
any other, merely that it is precise and that it provides
a reasonable starting point for discussion.

Specifically, in this note we attempt to convert the experimental results of
Draad and Nieuwstadt\cite{Dra,DKN} (henceforth DN) and Darbyshire and Mullin\cite{DaMu}
(henceforth DM) to $L^2$ amplitudes.
We conclude that in the $L^2$ framework, DN's published
exponent of $-1$ should be adjusted to between about $-2$ and $-1$,
and DM's published exponent of
between $-0.4$ and $-0.2$ should be adjusted to between
$-1.8$ and $-1.15$.\ \ One reason why these adjusted values are
expressed as ranges rather than
single numbers is that both sets of experiments introduce
perturbations by injection from the side of the pipe, and it is not known
exactly what perturbations these injections induce in the velocity
field within the pipe.  We emphasize that these ranges are rough, having
nothing like the two-digit precision suggested by
a number like $-1.15$.

Based on an asymptotic analysis of the Navier--Stokes
equations, Chapman\cite{Cha} has predicted the values $\gamma = -5/4$ for plane Couette
flow and $\gamma = -3/2$ for plane Poiseuille flow, and discussed the
relationship of these predictions to existing evidence from direct numerical
simulation of the Navier--Stokes equations (DNS)\cite{Hen,LHR}.\ \ (In the plane
Poiseuille case one
restricts attention to perturbations that avoid the Tollmien--Schlichting
instability.)  In work not yet written for publication, Chapman has extended
the prediction
$\gamma= -3/2$ also to pipe flow.  Thus if our adjusted exponents
for the DN and DM experiments are correct, there would appear to 
be reasonable agreement between two independent laboratories and a theoretical
calculation on an exponent for the pipe in the vicinity
\begin{equation}
\gamma \approx -{3\over 2}.
\label{eq2}
\end{equation}
This result would be consistent with the conjecture of Ref.~\cite{TTRD} that
$\gamma$ is strictly less than $-1$, an inequality also satisfied by most
of the low-dimensional ODE models of the Navier--Stokes
equations that have been published in the 1990s\cite{BaTr}.\ \ The apparent
convergence of various lines of evidence on the estimate (\ref{eq2}) looks promising,
but we urge that it not be taken as definitive or as numerically precise.
There are uncertainties at many points on
both the experimental and theoretical sides, and
no relevant data at all yet from  DNS simulations for the pipe.
Moreover, as we mention in Section~3, Chapman's asymptotic arguments
are based on pipe lengths much longer than those in the DN and DM experiments, and
for these pipes of finite lengths, somewhat less sensitivity to perturbations
may be expected.   We regard
(\ref{eq2}) as a rough working approximation.

\section{Nondimensionalization and $L^2$ amplitude}

One source of confusion about $\gamma$ has been the nondimensionalization of the
Navier--Stokes equations.  The $L^2$ definition of amplitude is formulated within
a particular choice of nondimensional variables, the standard one.  We
shall review this choice and explain why it can be a point of
confusion.

We are concerned with the idealized problem of laminar flow
through an infinite circular pipe.
The standard nondimensionalization
takes the pipe radius as the space scale and the centerline velocity as
the velocity scale.  Thus, after nondimensionalization, the radius
and velocity become
\begin{displaymath}
\hbox{radius} = 1, \quad \hbox{velocity} = 1.
\end{displaymath}
These choices imply that the nondimensional time scale is the convective
one, i.e., the time it takes the flow to travel downstream a distance
of one pipe radius:
\begin{displaymath}
\hbox{time to travel one pipe radius} = 1.
\end{displaymath}
Now there is also another time scale physically present in the
problem,
on which the effects of viscosity are felt.  In our
nondimensionalization this viscous time scale is $R$, the Reynolds number.
Thus we have
the following situation: a typical flow perturbation of small amplitude
and of spatial extent comparable to the pipe radius is convected
down the pipe at speed $O(1)$ for a time $O(R)$ and a distance $O(R)$
before the effects of viscosity become significant.

With these scales agreed upon, we imagine an initial value problem
in which at time $t=0$, the velocity field consists of the laminar solution
plus a divergence-free finite perturbation $u(0) = u(x,r,\theta,0)$.  The flow now
evolves according
to the Navier--Stokes equations, with the result that the initial perturbation
develops as a time-dependent divergence-free function
$u(t) = u(x,r,\theta,t)$.\ \ At any time
$t$, we measure the amplitude of $u$ in an $L^2$ fashion:
\begin{equation}
\|u(t)\| \;=\; \left(\;
\int_{-\infty}^\infty\int_0^1\int_0^{2\pi} u(x,r,\theta,t)^2 \kern 2pt
d\theta \kern 1.4pt r \kern .6pt dr
\kern 1pt dx \right)^{1/2}.
\label{eq3}
\end{equation}
Thus $\|u(t)\|$ is the root-mean-square velocity perturbation over the whole pipe.

We now return to the matter of why these formulations may sometimes
be confusing.
In most laboratory experiments, and certainly in DM and DN, the
Reynolds number $R$ is
controlled by varying the speed of the flow,
not the viscosity.  On the other hand in nondimensional units the speed of
the flow is always $1$, and other velocities are defined as ratios to this one.
Thus to convert from laboratory to nondimensional units we must
\begin{equation}
\hbox{multiply time measured in seconds by } R
\label{eq4}
\end{equation}
and
\begin{equation}
\hbox{divide velocity measured in meters/second by } R
\label{eq5}
\end{equation}
(as well as $R$-independent scalings by pipe diameter divided
by kinematic viscosity).
In particular, the nondimensionalized $O(R^{\kern 1pt \gamma})$ and $O(R^{-3/2})$ formulas of
(\ref{eq1}) and (\ref{eq2}) would appear as
$O(R^{\kern 1pt\gamma+1})$ and $O(R^{-1/2})$ in
laboratory units.  Thus (\ref{eq2}) can be paraphrased by 
the statement that if you double the speed of flow of water
through an infinitely long pipe, the minimal velocity perturbation needed to
excite transition becomes
smaller in meters/second by a factor of about $\sqrt 2$
and
smaller relative to the flow speed by a factor of about $2\sqrt 2$.

\section{Asymptotic estimates of Chapman}

Chapman's paper\cite{Cha} estimates $\gamma$ for channel flows
by asymptotic analysis of the Navier--Stokes equations.  The exponent
$\gamma = -3/2$ is obtained for plane Poiseuille flow, and though this
has not yet been written for publication, the same exponent results from
an analogous analysis of pipe flow.  Chapman uses the 
$L^2$ definition of amplitude as described above, except that he assumes
a periodic flow perturbation and defines amplitude by an integral over
one period.  We believe that
this does not affect the final result, so that his conclusion can be
fairly summarized by (\ref{eq2}).

We shall say nothing of the arguments of Ref.\ \cite{Cha} except to note that
they are based on the specific initial condition that appears to be
most effective at exciting transition, a streamwise vortex
plus small non-streamwise components.  If the analysis is correct, the
threshold amplitude for such perturbations to excite transition will
scale as $R^{-3/2}$ as $R\to\infty$.\ \ In principle $-3/2$ is thus
a proposed upper bound for $\gamma$ in the sense that there is
the possibility
that some other initial configuration might be found that would
excite transition more effectively.

It must be noted, however, that Chapman's analysis is based on
flow structures that evolve on a time scale $O(R)$, during which they
move a distance $O(R)$ and stretch a distance $O(R)$. 
For the mechanisms involved to come fully into play, a pipe would
have to have length at least $O(R)$, i.e., $O(R)$ pipe diameters.  This
exceeds the actual pipe lengths of experiments, which are in the
hundreds, not thousands or tens of thousands.  Thus
an exponent as low as the theoretical value
of $-3/2$ should not necessarily
be observable in any existing pipe experiment.
(Conversely, Chapman also identifies other finite-$R$ effects that
act in the opposite direction, effects which make
exponents $\gamma$ estimated from data with $R < 10^4$ more negative
than the asymptotic values for $R\to\infty$.)

\section{Pipe experiments of Draad and Nieuwstadt}

The DN experiments in the 36m pipe at the Delft
University of Technology are described in detail in Ref.\ \cite{Dra}.\ \ In these
experiments, a disturbance is introduced into the laminar flow through
a set of slits in the side of the pipe.
These are pumped in an oscillatory fashion so that water is injected
and extracted sinusoidally at a controllable frequency and amplitude.
As a measure of disturbance amplitude, DN
take injection velocity nondimensionalized according to (\ref{eq5}), i.e.,
divided by the flow velocity in the pipe.  Their experiments lead
to the following estimates, summarized for example in Figure~6.8
on p.~140 of Ref.\ \cite{Dra}:
\begin{equation}
\hbox{DN as published:} \quad \gamma \approx \cases{
-2/3 & \hbox{for long wavelengths,}\cr
 -1  & \hbox{for short wavelengths.}\cr}
\label{eq6}
\end{equation}
``Long'' and ``short'' wavelengths are defined in the usual nondimensional space
scale, i.e., relative to the pipe radius.
For simplicity, since our interest is in smallest perturbations that
may excite turbulence, from now on we shall consider just the value
$-1$ reported by DN for short wavelengths.

Several matters arise in the attempt to
convert the DN result to the amplitude measure (\ref{eq3}).
The most obvious is the fact that since the DN perturbations are periodic in time, they
have infinite $L^2$-amplitude.  For the conversion to (\ref{eq3}) we must
guess how short a finite-length perturbation might have led to approximately
the same observations.  If a perturbation of length $O(1)$ (i.e., of length
independent of $R$) would suffice, then the exponent $\gamma\approx -1$ can be
taken at face value.  On the other hand one might
also imagine that perturbations of
length and time scale $O(R)$ (i.e., laboratory time $O(1)$ as measured in seconds)
would be needed.  In this case the disturbance amplitudes must be multiplied
by $O(R^{1/2})$ for conversion to $L^2$ amplitude because of the square root in
(\ref{eq3}), so that the estimate $\gamma\approx -1$ should be increased by $1/2$ to $-1/2$.

Thus at this stage of the discussion it would
appear that the DN minimal exponent $-1$
corresponds in the amplitude measure (\ref{eq3})
to a figure in the range $-1$ to $-1/2$.

It appears to us that there is also a second adjustment that 
should be applied to cast these results
in terms of the $L^2$ amplitude (\ref{eq3}).
DN measure disturbance amplitude by velocity in the
injection and extraction slits.
However, the velocity of the water in the slits is
not proportional to the velocity of the perturbation it induces in the pipe.
The reason is that as the flow speed in the pipe increases with $R$,
a proportionally greater volume of water is disturbed by the injection
and extraction, implying that the pointwise velocity disturbance
shrinks.

{\em Penetration scenario.}\ \ Suppose that
for any flow speed $R$, the injected perturbation penetrates 
approximately the full width of the
pipe.  Then because the amount of fluid into which it is injected scales as
$O(R)$, the velocity amplitude
reduces pointwise by $O(R)$.\ \ The
exponents $-1$ to $-1/2$ would need
to be decreased by $1$, giving the range from $-2$ to $-3/2$.

{\em Non-penetration scenario.}\ \ On the other hand,
it is not obvious that injected perturbations penetrate the pipe effectively,
and the other extreme scenario would
seem to be that the injected perturbation affects a region near the pipe wall
of width $O(R^{-1})$.  In this case, the perturbation is
distributed over a volume $O(R^{-1})\times O(R) = O(1)$, i.e., a volume independent
of $R$.\ \ The pointwise velocity amplitude will accordingly be independent of $R$
in that region.  At the same time, the fraction of the pipe filled
by the velocity perturbation is now not $O(1)$ but $O(R^{-1})$, implying an
$L^2$ correction factor of $O(R^{-1/2})$.\ \ Thus according to
the $L^2$ definition of amplitude, the exponents
$-1$ to $-1/2$ would need to be
decreased by $1/2$, giving the range from $-3/2$ to $-1$.

Our discussion has raised two physical questions (whose answers may be related).
Rather than attempt to resolve them on the basis of meagre
evidence, we summarize our current understanding of the DN observations for short
wavelengths by the range
\begin{displaymath}
\hbox{DN adjusted to $L^2$ amplitudes:}\quad {}-2 \le \gamma \le 1.
\end{displaymath}
In principle these are estimated upper bounds for $\gamma$, as it is always
possible that the perturbations actually injected are not maximally
efficient in exciting turbulence.

\section{Pipe experiments of Darbyshire and Mullin}

The DM experiments, described in Ref.\ \cite{DaMu}, were carried out in a 3.8m pipe
at Oxford University.  (The equipment subsequently moved with Mullin to the University
of Manchester, where a new 17m pipe has recently been built based on the same design.)
These experiments differ in many ways from those of
DN, the most fundamental one being that water is sucked out of the
pipe at fixed speed rather than pushed into it at fixed pressure.
Another important difference is that
whereas the DN perturbation is periodic, the DM perturbation is injected
just once.

The DM paper does not propose a value for $\gamma$ except to suggest
that it seems to be just slightly less than $0$.\ \ Based on the plots in
Ref.\ \cite{DaMu}, a rough estimate would seem to be
\begin{equation}
\hbox{DM as published:}
\quad {}-0.4 \le \gamma \le -0.2. 
\label{eq7}
\end{equation}
However, the disturbance amplitudes reported by DM are not
normalized by the velocity in the pipe.  Introducing the adjustment
(\ref{eq5}) gives
\begin{displaymath}
\hbox{DM after nondimensionalization:}
\quad {}-1.4 \le \gamma \le -1.2.
\end{displaymath}

Now the more difficult question arises, as in the DN case, of what
further adjustment may be needed because
of the complex relationship between the flow velocity in the slits and the perturbation
velocities induced in the pipe.
As with DN, the DM flow perturbations are measured
in the slits, not in the pipe.
The form of the injections even in the slits is complicated by
the geometry of a drive mechanism mounted on a rotating plate.
In adjusting what they call the amplitude $A$, DM increase the injected volume
in proportion to $A$, the injection time approximately in proportion to $A^{1/2}$,
and the maximum injection velocity approximately in proportion to
$A^{1/2}$.\ \ Within the pipe, this will produce a velocity perturbation
extending a distance on the order of $A^{1/2} R$.

{\em Penetration scenario.}\ \ Suppose the perturbation
penetrates a distance $O(1)$ into the pipe.
Then the volume of the effective disturbance is $O(A^{1/2} R)$
and its pointwise amplitude is $O(A^{1/2}R^{-1})$,
giving an $L^2$ amplitude $O(A^{3/4}R^{-1/2})$.

{\em Non-penetration scenario.}\ \ Suppose the perturbation
penetrates only a distance $O(R^{-1})$ into the pipe.  Then the volume
of the effective disturbance is $O(A^{1/2})$ and its pointwise amplitude is
$O(A^{1/2})$,
giving an $L^2$ amplitude of $O(A^{3/4})$.

We conclude that two further adjustments of the DM results are needed to
convert them to $L^2$ amplitudes.
First, because what DM call $A$ becomes $O(A^{3/4})$ in the $L^2$ measure,
the numbers $-0.4$ and $-0.2$ should be multiplied by $3/4$, becoming
$-0.3$ and $-0.15$.
Second and more important, the final numbers
obtained should be reduced by between $0$ and $1/2$.\ \ Putting all these
adjustments together gives:
\begin{displaymath}
\hbox{DM adjusted to $L^2$ amplitudes:}\quad -1.8 \le \gamma \le -1.15.
\end{displaymath}
We emphasize once more that our arguments and measurements are not
as precise as these numbers may suggest.

In principle these are again estimated upper bounds for $\gamma$, as it is again
possible that the perturbations actually injected are not maximally
efficient.  In particular, it might be possible to excite turbulence
more efficiently by disturbances shaped to have a streamwise length
with a different dependence on $R$ and on amplitude.

\section{Discussion}

The existing experimental and theoretical literature on
threshold exponents for transition in a pipe is based on inconsistent
definitions of amplitudes, so the published results are not comparable.
Here we have attempted to reformulate some of these results in a
matter consistent enough for a meaningful comparison.  There are numerous
uncertainties in this process, including
spatial vs.\ temporal growth of disturbances, solitary vs.\ periodic
disturbances, form of the velocity field perturbation effectively introduced by injection,
non-``optimality'' of experimentally injected disturbances from the point
of view of exciting transition, the effects of finite pipe length, and, of
course, experimental error.
For all these reasons, no decisive conclusion can be drawn from our comparison.
The tentative conclusion we draw is that
the reformulated experimental (Draad and Nieuwstadt;
Darbyshire and Mullin) and theoretical (Chapman) results
appear to agree upon a critical exponent roughly
in the range $\gamma = -3/2\pm 0.3$, if the centerline velocity is
nondimensionalized to~$1$.

\acknowledgements

The research of AM, TM, and LNT was supported by
the Engineering and Physical Sciences
Research Council of the UK under Grant GR/M30890.
The work of TM was also supported by the Leverhulme Trust.

\end{document}